\documentclass[twocolumn,prb,amssymb,showpacs]{revtex4}
\usepackage{graphicx}

\begin{document}

\title{Diffusion and ballistic contributions of the interaction correction to
the conductivity of a two-dimensional electron gas}

\author{G.~M.~Minkov}
\author{A.~A.~Sherstobitov}
\affiliation{Institute of Metal Physics RAS, 620219 Ekaterinburg,
Russia}

\author{A.~V.~Germanenko}
\author{O.~E.~Rut}
\author{V.~A.~Larionova}
\affiliation{Institute of Physics and Applied Mathematics, Ural
State University, 620083 Ekaterinburg, Russia}

\author{A.~K.~Bakarov}
\affiliation{ Institute of Semiconductor Physics,  630090
Novosibirsk, Russia}

\author{B.~N.~Zvonkov}
\affiliation{Physical-Technical Research Institute, University of
Nizhni Novgorod,  603600 Nizhni Novgorod, Russia}

\date{\today}

\begin{abstract}
The results of an experimental study of interaction quantum
correction to the conductivity of two-dimensional electron gas in
A$_3$B$_5$ semiconductor quantum well heterostructures are presented
for a wide range of $T\tau$-parameter ($T\tau\simeq 0.03-0.8$),
where $\tau$ is the transport relaxation time. A comprehensive
analysis of the magnetic field and temperature dependences of the
resistivity and the conductivity tensor components allows us to
separate the ballistic and diffusion parts of the correction. It is
shown that the ballistic part renormalizes in the main the electron
mobility, whereas the diffusion part contributes to the diagonal and
does not to the off-diagonal component of the conductivity tensor.
We have experimentally found the values of the Fermi-liquid
parameters describing the electron-electron contribution to the
transport coefficients, which are found in a good agreement with the
theoretical results.
\end{abstract}

\pacs{73.20.Fz, 73.61.Ey}

\maketitle
\section{Introduction}
The temperature and magnetic field dependences of the resistivity of
the degenerated two dimensional (2D) gas at low temperatures are
determined by the quantum corrections to the conductivity. They are
the weak localization (WL) or interference correction and the
correction caused by the electron-electron ({\it e-e}) interaction.
The WL correction to the conductivity in the absence of the spin
relaxation is negative and logarithmically increases in absolute
value with decreasing temperature. The interaction correction for
low $r_s$-values and within the diffusion regime, $T\tau\ll 1$
(where $r_s$ is the gas parameter,  and $\tau$ is the transport
relaxation time, hereafter we set $\hbar=1$, $k_B=1$) is, as a rule,
negative and also increases in absolute value with the temperature
decrease. However, the detailed theoretical analysis of the
interaction correction to the conductivity for the intermediate
($T\tau\simeq 1$) and ballistic ($T\tau\gg 1$) regimes carried out
in Refs.~\onlinecite{Zala01,Gor03,Gor04} shows that this correction
can result in different sign of the $T$-dependence of the
resistivity. It is dictated by the value of the Fermi-liquid
constant (see Figs.~7 and 8 in Ref.~\onlinecite{Zala01}). Just this
fact comes to special attention to the interaction correction in
ballistic regime  because it can explain the metallic-like
temperature dependence of the conductivity
\cite{Pudalov03,Li03,Prosk02,Gal04} observed in  some 2D structures.

The {\it e-e} correction in the diffusion regime can be easily
separated experimentally because it contributes only to diagonal
components of the conductivity tensor and does not to off-diagonal
one. It is more difficult to extract experimentally the correction
in the intermediate and the ballistic regimes  because the theories
does not divide the interaction correction into the diffusion and
ballistic parts and does not predict any specific features of this
correction. Besides, some classical mechanisms such as temperature
dependent disorder,\cite{Alt01} classical magnetoresistance due to
scattering by rigid scatterers\cite{Dmit01} can complicate the
situation.\cite{Li03,Gal04} In the  ballistic regime, $T\tau\gg 1$,
for the white noise disorder and classically  low magnetic field,
$\mu B\ll 1$, where $\mu$ is the mobility, the
theory\cite{Zala01,Zala01-1} predicts that the {\it e-e} interaction
contributes both to $\sigma_{xx}$ and $\sigma_{xy}$ in such a way
that it does not influence the Hall coefficient, $R_H$. This means
that the {\it e-e} interaction in this regime reduces to a
renormalization of the transport relaxation time. The same result
was obtained for the long-range and mixed disorder  at high magnetic
field, $\mu B\gg 1$, in Refs.~\onlinecite{Gor03} and
\onlinecite{Gor04}.

In this  paper we  systematically study the {\it e-e} interaction
correction to the conductivity of $n$-type
Al$_x$Ga$_{1-x}$As/GaAs/Al$_x$Ga$_{1-x}$As and
GaAs/In$_x$Ga$_{1-x}$As/GaAs  quantum wells.  The comprehensive
analysis of the  data within wide $T\tau$ range ($T\tau=0.03-0.8$)
and classically strong magnetic field shows that the interaction
correction to the conductivity can be divided into two parts. The
first part  contributes to $\sigma_{xx}$ only, it is proportional to
$\ln[1/(T\tau)+1]$ within whole $T\tau$ range (we refer to this part
as ``diffusion part''), while the second one reduces to
renormalization of the transport relaxation time and is proportional
to $T\tau$ (this part is termed as the ``ballistic part'').

\section{Theoretical background}
\label{sec:TB}

The conductivity of a system at zero magnetic field is given by
\begin{eqnarray}
{\sigma=\sigma_0+\delta \sigma^{ee}+\delta\sigma^{WL}}
\label{eq01}.
\end{eqnarray}
Here,  $\sigma_0=en\mu$ with $n$ as the electron density, is the
Drude conductivity, $\sigma^{WL}$ and $\delta \sigma^{ee}$ stand for
the weak-localization and interaction quantum correction,
respectively. The weak-localization correction looks as follows
\begin{eqnarray}
\frac{\delta\sigma^{WL}}{G_0}&=& - \ln{\left(1+\frac{\tau_\phi}
{\tau}\right)} \nonumber \\
&+&\frac{1}{1+2\tau_\phi/\tau}\ln{\left(1+\frac{\tau_\phi}
{\tau}\right)}+ \frac{\ln{2}}{1+\tau/2\tau_\phi}, \label{eq011}
\end{eqnarray}
where $G_0=e^2/(2\pi^2\hbar)\simeq 1.23\times
10^{-5}$~$\Omega^{-1}$, $\tau_\phi$ is the phase relaxation time,
and the second and third terms take into account
non-backscattering processes.\cite{Dmit97} The interaction
correction was calculated  in Refs.~\onlinecite{Zala01} and
\onlinecite{Gor04}, and for a white noise disorder and wide range
of $T\tau$-values  is given by:
\begin{eqnarray}
{\delta \sigma^{ee}\over G_0}&=&2\pi
T\tau\left[1-\frac{3}{8}f(T\tau)+\frac{3\widetilde{F}_0^\sigma}{1+\widetilde{F}_0^\sigma}\right.\nonumber \\
&\times&
\left.\left(1-\frac{3}{8}t(T\tau,\widetilde{F}_0^\sigma)\right)\right]
\nonumber \\
 &-& \left[1+3\left(1-\frac{\ln(1+F_0^\sigma)}{F_0^\sigma}\right)\right]\ln{\frac{E_F}{T}}
\label{eq02},
\end{eqnarray}
The functions  $f(T\tau)$ and $t(T\tau,\widetilde{F}_0^\sigma)$ are
given in Ref.~\onlinecite{Zala01}. In contrast to Eq. (2.16c) of
this paper we have explicitly written the different Fermi-liquid
constants in the first and second terms of Eq.~(\ref{eq02}) (see
last paragraph in pages 5  of the same paper). When the
$T\tau$-value is low enough, the temperature dependence of
$\delta\sigma^{ee}$ is controlled by the second term in
Eq.~(\ref{eq02}) and it is logarithmic. For the high $T\tau$-value,
the first term in Eq.~(\ref{eq02}) becomes dominant, because the
functions $f(T\tau)$ and $t(T\tau,\widetilde{F}_0^\sigma)$  go to
zero when $T\tau\to\infty$. In this limit, the conductivity changes
with the temperature linearly. Noteworthy is the argument of the
logarithm in Eq.~(\ref{eq02}), which is written by the authors as
$E_F/T$ instead of the usual $1/T\tau$.\cite{AA85}

Unfortunately,  it is not sufficient for the reliable determination
of $\delta\sigma^{ee}$ to know the behavior of the interaction
correction in the absence of magnetic field. The reason is the other
temperature dependent scattering mechanisms, for instance, phonon
scattering or the temperature-dependent disorder,\cite{Alt01} which
can be presented in real systems and can mask the effect under
consideration. Investigations in the presence of  magnetic field are
much more informative from this point of view.

In a magnetic field the conductivity tensor can be written as
\begin{eqnarray}
\sigma_{xx}&=&\frac{en\mu}{1+\mu^2 B^2}+\delta\sigma_{xx}^d+\delta\sigma_{xx}^b,  \label{eq03} \\
\sigma_{xy}&=&\frac{en\mu^2 B}{1+\mu^2 B^2}+\delta\sigma_{xy}^b,
\label{eq04}
\end{eqnarray}
where $\delta\sigma_{xx}^d$ and $\delta\sigma_{xx}^b$ are the
diffusion and ballistic parts of the interaction correction. It is
important to mention here that the diffusion part of the
electron-electron interaction contributes to $\sigma_{xx}$ only and
does not to $\sigma_{xy}$.\cite{AA85} This is a key feature of the
diffusion correction, which allows one to determine its value
experimentally. The diffusion correction $\delta\sigma_{xx}^d$
logarithmically depends on the temperature and does not depend on
the magnetic field (the latter is true if the Zeeman splitting is
less than $T$). It is usually written as\cite{AA85,Fin,Cast}
\begin{eqnarray}
{\delta \sigma_{xx}^{d}(T)\over G_0}&=&-\left[1+3\left(1-
\frac{\ln(1+F_0^\sigma)}{F_0^\sigma}\right)\right]\ln{\frac{1}{T\tau}}\nonumber
\\
&\equiv&K_{ee}\ln{T\tau}, \label{eq05}
\end{eqnarray}
where  the first term in square brackets  is the exchange or the
Fock contribution while the second one is the Hartree contribution
(the triplet channel). Comparing Eq.~(\ref{eq02}) with
Eq.~(\ref{eq05}) one can see that the arguments in logarithms in
these expressions are different and distinguished by a factor
$E_F\tau=k_Fl/2$. Thus, the question whether the argument in the
logarithm in Eq.~(\ref{eq05}) is $E_F/T$ or $1/(T\tau)$ is open.

As for the ballistic contributions, the situation is more
complicated. The ballistics contribute both to $\sigma_{xx}$ and to
$\sigma_{xy}$, and, in general case, $\delta\sigma_{xx}^b$ and
$\delta\sigma_{xy}^b$ depend both on the magnetic field and
temperature. For the low magnetic field, $B\ll 1/\mu $, and
white-noise disorder the corrections  to the conductivity tensor
components  were calculated in Refs.~\onlinecite{Zala01,Zala01-1},
while for the high magnetic field and smooth or mixed disorder it
was done in Ref.~\onlinecite{Gor04}. The results for all the cases
are different, however, the analysis shows  that in the limiting
case $T\tau\gg1$ the interaction correction universally reduces to a
renormalization of the transport relaxation time. It is physically
understandable because  the interaction correction in this regime
can be considered as a result of elastic scattering of an electron
by the temperature-dependent  self-consistent potential created by
all the other electrons.\cite{Zala01} It is reasonable to generalize
this result and to assume that the ballistic correction reduces in
the most part to the renormalization of the transport relaxation
time, i.e., to the renormalization of the mobility for any values of
$T\tau$-parameter. Then, Eqs.~(\ref{eq03}) and (\ref{eq04}) can be
rewritten as follows
\begin{eqnarray}
\sigma_{xx}&\simeq&\frac{en\mu'}{1+\mu'^2 B^2}+\delta\sigma_{xx}^d,  \label{eq06} \\
\sigma_{xy}&\simeq&\frac{en\mu'^2 B}{1+\mu'^2 B^2}, \label{eq07}
\end{eqnarray}
where $\mu'=\mu+\delta\mu$ is the mobility renormalized by the
ballistics. Such an assumption is in accordance with the results for
different limiting cases obtained the papers cited above. In
particular, within these frameworks one obtains the logarithmic
behavior of $\sigma(B=0)$, $\sigma_{xx}$ and the Hall coefficient at
low temperatures, $T\ll 1/\tau$, and the vanishing of the
interaction correction to the Hall coefficient at
$T\tau\to\infty$.\cite{Zala01,Gor04}

In what follows we will show that such a model for the ballistic
correction well describes the experimental data at low and
intermediate  temperatures up to $T\tau\simeq 1$.

Below we will name that part of the interaction correction which
contributes to  $\sigma_{xx}(B)$ but does not to $\sigma_{xy}(B)$ as
``the diffusion correction'' because just the same property has the
{\it e-e} correction in the diffusive regime. The part of
interaction correction which renormalizes the mobility $en
\delta\mu$ we will name as ``the ballistic correction''.

\section{Experiment}

We study the interaction correction to the conductivity in
heterostructures of two types. The first one is the
Al$_x$Ga$_{1-x}$As/GaAs/Al$_x$Ga$_{1-x}$As quantum well
heterostructure grown by MBE on semiinsulator GaAs substrate. It
consists of $250$~nm-thick undoped GaAs buffer layer, a $50$~nm
Al$_{0.3}$Ga$_{0.7}$As barrier, Si $\delta$-layer, a $6$~nm spacer
of undoped Al$_{0.3}$Ga$_{0.7}$As, a $8$~nm GaAs well, a $6$~nm
spacer of undoped Al$_{0.3}$Ga$_{0.7}$As, a Si $\delta$-layer, a
$50$~nm Al$_{0.3}$Ga$_{0.7}$As barrier, and $200$~nm cap layer of
undoped GaAs. The second structure is GaAs/In$_x$Ga$_{1-x}$As/GaAs
structure. It consists of a $200$~nm-thick undoped GaAs buffer
layer, Si $\delta$-layer, a 9~nm spacer of undoped GaAs, a 8~nm
In$_{0.2}$Ga$_{0.8}$As well, a 9~nm spacer of undoped GaAs, a Si
$\delta$-layer, and 200~nm cap layer of undoped GaAs. The samples
were mesa etched into standard Hall bars and then an Al gate
electrode was deposited by thermal evaporation onto the cap layer
of the first structure through a mask. Varying the gate voltage
$V_g$ from $1$~V to $-4$~V we decreased the electron density in
the quantum well from $1.7\times 10^{12}$ cm$^{-2}$ to $7\times
10^{11}$~cm$^{-2}$. The electron density in the second structure
was controlled through the illumination due to  the persistent
photoconductivity effect. The measurements were performed after
the illumination of the different intensity and duration that
allowed us to change the electron density within the range from
$4.3\times 10^{11}$ cm$^{-2}$ to $7\times 10^{11}$~cm$^{-2}$.
Analysis of the Shubnikov-de Haas oscillations in the first
structure shows that the second subband starts to be occupied at
$V_g\simeq 0$. In order to prevent the multiband effects we will
analyze the results obtained for $V_g\leqslant -1$ V, when the
excited subbands lay far above the Fermi level and are practically
empty for the actual temperature range. (So for $V_g=-1$ V, the
Fermi level lies about $5$~meV below the bottom of the second
subband and the estimated value of the electron density in this
subband is about $10^8$~cm$^{-2}$ for $T=10$~K. For the lower
temperatures and gate voltages this quantity is far less).
\begin{table}[b]
\caption{The parameters of structures investigated} \label{tab1}
\begin{ruledtabular}
\begin{tabular}{ccccc}
Structure & $V_g$ (V) &$n$ ($10^{12}$ cm$^{-2}$) & $\mu$ (cm$^2$/V~s)  & $B_{tr}$ (mT)\\
 \colrule
 T1520          &$-1.0$   &$1.303$   &$14470$ &4.4\\
              &$-2.5$   &$0.967$   &$8950$  &15.7\\
               &$-3.7$   &$0.715$   &$4925$ &70.0\\
 3510\footnotemark[1]          &   &$0.44$   &$19300$  &25.9\\
                &   &$0.56$   &$16000$ &8.5\\
                &   &$0.7$   &$10400$  &4.7\\
 \end{tabular}
\end{ruledtabular}
\footnotetext[1]{The electron density in this structure is changed
via the illumination.}
\end{table}

\begin{figure}
\includegraphics[width=\linewidth,clip=true]{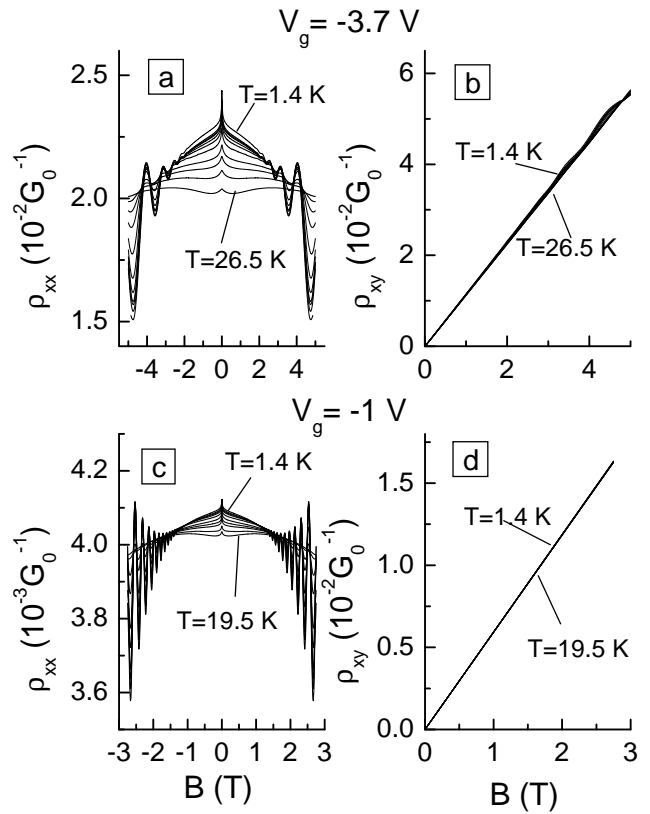}
\caption{The magnetic field dependences of $\rho_{xx}$ (a,c) and
$\rho_{xy}$ (b,d) measured for different temperatures and
$V_g=-3.7$~V (a,b) and $-1$~V (c,d). Structure T1520. The
temperature for curves are: $1.4$, $2.0$, $2.5$, $3.0$, $4.2$,
$6.2$, $9.0$, $13.0$, $16.5$, $20.0$, $26.5$~K (a,b) and $1.4$,
$2.0$, $2.5$, $3.0$, $4.2$, $6.0$, $8.5$, $11.0$, $16.0$, $19.5$~K
(c,d).}
 \label{F1}
\end{figure}

We  measured carefully the low and high magnetic field longitudinal
($\rho_{xx}$) and transverse ($\rho_{xy}$) magnetoresistance in a
magnetic field up to $5$~T within the temperature range from $0.4$
to $25$~K. The detailed measurements were performed for three gate
voltages for the first structure and after three illumination flux
for the second one. The main  parameters of the structures are given
in Table~\ref{tab1}, in which  $B_{tr}$ stands for the transport
magnetic field defined as $B_{tr}=\hbar/2e l^2$, where $l$ is the
mean free path. The electron density and mobility have been found
from the fit of $\sigma_{xy}$-vs-$B$ plots, the value of $\mu$ given
in the table relates to $T=0$ and is obtained by a linear
extrapolation of the experimental dependence $\mu(T)$  (see text
below).

To elucidate the role of the ballistic contribution of the {\it e-e}
interaction we will consider in parallel the experimental data
obtained for the structure T1520 for two limiting gate voltages:
$V_g=-3.7$ V, when the diffusion contribution is dominant, and
$V_g=-1$ V, when the ballistics become important.

Fig.~\ref{F1} shows the magnetic field dependences of $\rho_{xx}$
and $\rho_{xy}$ measured at different temperatures. One can see
from Figs.~\ref{F1}(a) and \ref{F1}(c) that following the sharp
magnetoresistance in low magnetic field [evident  at $B\lesssim
0.05$~T in Fig.~\ref{F1}(a) and  at $B\lesssim 0.02$~T in
Fig.~\ref{F1}(c)], which results from the suppression of the
interference quantum correction,\cite{Min05} the parabolic-like
negative magnetoresistance against the background of the
Shubnikov-de Haas oscillations is observed. The parabolic-like
behavior of $\rho_{xx}$ weakens with the increasing temperature
transforming to nonmonotonic one at $T\gtrsim 20$~K. The
transverse magnetoresistance $\rho_{xy}$ slightly decreases with
increasing temperature [Figs.~\ref{F1}(b) and \ref{F1}(d)]. On the
first sight the behavior of the resistivity tensor components is
identical for both gate voltages. However, some quantitative
difference occurs as we will show below.

\begin{figure}
\includegraphics[width=\linewidth,clip=true]{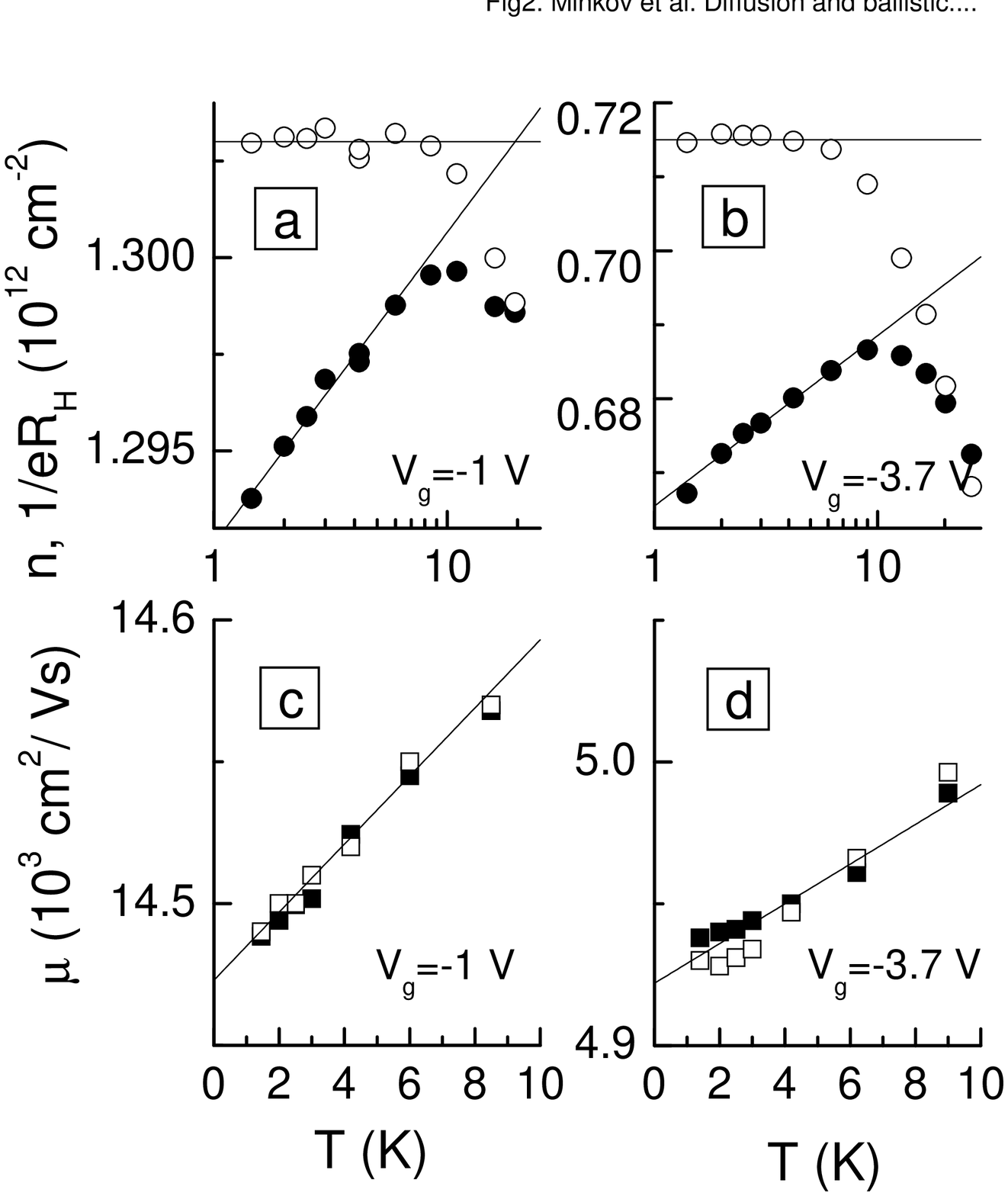}
\caption{The temperature dependences of $1/eR_H$ ($\bullet$), $n$
($\circ$) (a, b) and $\mu$ (c, d) for structure  T1520, measured for
$V_g=-1$~V (a,c) and $-3.7$~V (b,d). Symbols are the experimental
data. The values of electron density, $n$, and mobility $\mu$ shown
by open symbols have been obtained from the fitting of
$\sigma_{xy}$-vs-$B$ experimental plots. The solid symbols for $\mu$
are  obtained from the fit of $\sigma_{xx}$-vs-$B$ curves. Lines in
(a) and (b) are provided as a guide for the eye. Straight lines in
(c) and (d) are drawn through the experimental points and show the
extrapolation to $T=0$. }
 \label{F2}
\end{figure}

Let us begin our analysis with the temperature dependence of the
Hall coefficient, $R_H=\rho_{xy}/B$. Its value has been found from a
linear  interpolation of the $\rho_{xy}$-vs-$B$ dependence made in
the magnetic field range $-1/\mu\ldots 1/\mu$.\footnote{ The
detailed studies of the low field behavior of $\rho_{xy}$ reveals
some nonlinearity of the $\rho_{xy}$-vs-$B$ dependence  at
$B<(2-3)B_{tr}$. Just in this magnetic field range the WL correction
determines the magnetic field dependence of $\rho_{xx}$. Such a
behavior was already reported in Ref.~\onlinecite{Pool81,Newson87}
and, we believe, exists in the most of the 2D structures.  Probably,
experimentalists do not like to publish the results of such a kind,
because the theory does not predict the magnetic field dependence of
the Hall coefficient due to the weak localization.} The results are
presented  as $1/(eR_H)$-vs-$T$ plots in Figs.~\ref{F2}(a) and
\ref{F2}(b) by the solid symbols. One can see that the quantity
$1/(eR_H)$ increases logarithmically with the increasing
temperature, while the temperature remains less than $8-10$ K.
Namely such a behavior is predicted by the theories for the
diffusion regime (see Section \ref{sec:TB}). However, at higher
temperature the value of $1/(eR_H)$ surprisingly starts to fall.

To understand, whether the high-temperature behavior of $1/(eR_H)$
results from the lowering  of the electron density  with
$T$-increase or it is a peculiarity of the {\it e-e} interaction,
the Shubnikov-de Haas oscillations have been analyzed. It turns
unfortunately out that it is impossible to find the electron density
from the period of the oscillations at high temperature with the
accuracy required (as seen from Figs.~\ref{F2}(a) and \ref{F2}(b)
the fall does not exceed $1-2$~\% in magnitude in our temperature
range).

\begin{figure}
\includegraphics[width=\linewidth,clip=true]{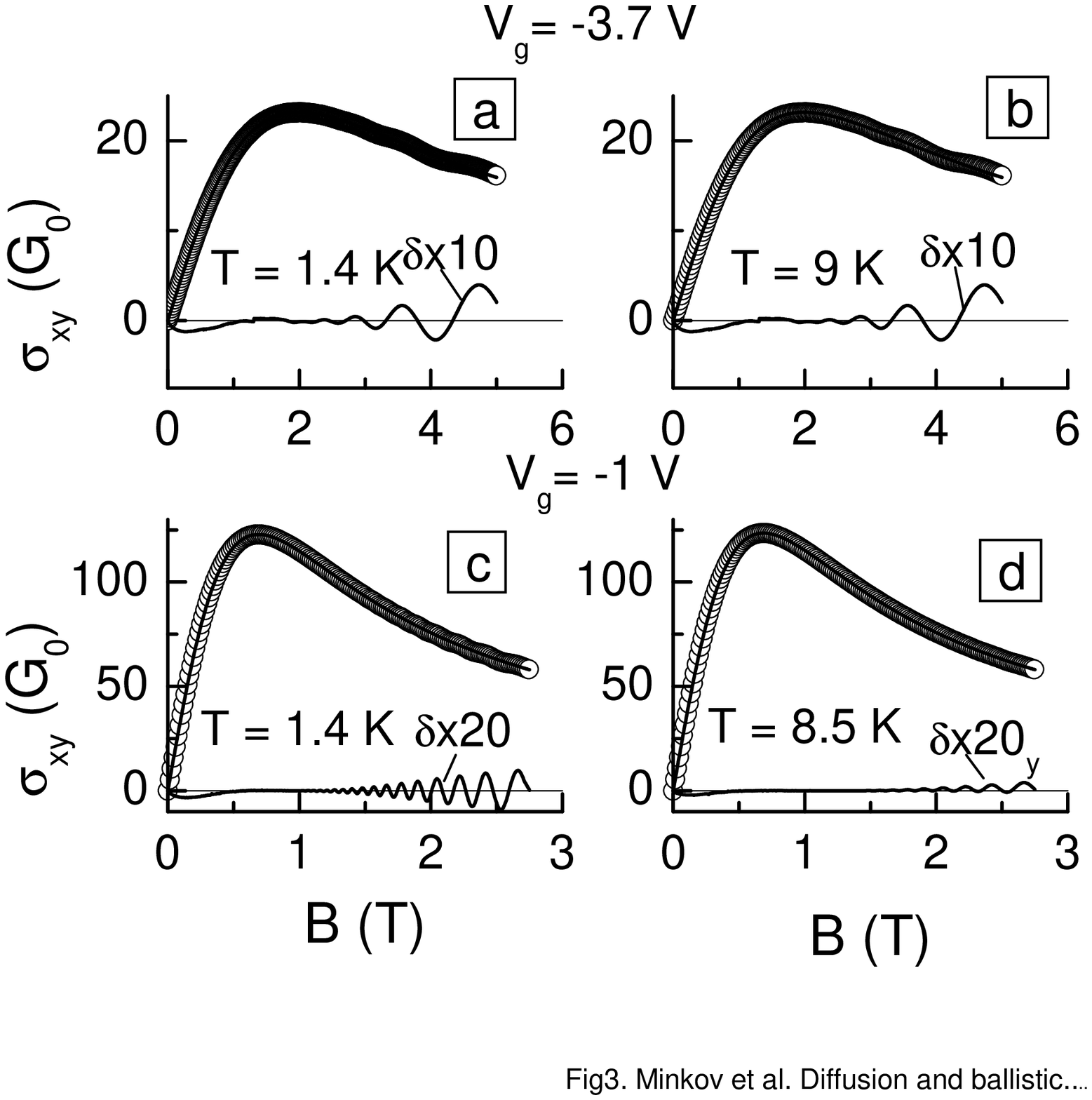}
\caption{The magnetic field dependences of $\sigma_{xy}$ measured
for different temperatures for $V_g=-3.7$~V (a,b) and $V_g=-1$~V
(c,d), structure T1520. Symbols are the experimental data, solid
curves are the fit by Eq.~(\ref{eq07}) with $n$ and $\mu'$ as
fitting parameters. Curves labeled as $\delta\times 10$ (a,b) and
$\delta\times 20$ (c,d) are the difference between experimental and
fitting curves multiplied by a factor $10$ or $20$, respectively.}
 \label{F3}
\end{figure}

Another way to find the electron density  is the analysis of the
magnetic field dependence of $\sigma_{xy}$ because it is unaffected
by the {\it e-e} interaction in the diffusion regime. The fit of the
experimental data by Eq.~(\ref{eq07})  with $n$ and $\mu'$ as the
fitting parameters gives a very reproducible result. Fig.~\ref{F3}
shows the result of such a fit  made for magnetic field range from
$20\,B_{tr}$ to $B=1.5/\mu$.  A nice coincidence with the
experimental $\sigma_{xy}$-plots is evident. The value of $n$
obtained for different temperatures by this way is presented in
Figs.~\ref{F2}(a) and \ref{F2}(b) by open symbols.
As seen it is constant at low temperature and decreases at
$T>10$~K, coinciding  practically with $1/(eR_H)$-value at
$T\gtrsim 15$~K. Note, such a  behavior of $1/(eR_H)$ and $n$ with
temperature holds when the wider fitting interval of the magnetic
field is used. Thus, we believe that the fall of $1/(eR_H)$ and
$n$ evident at $T>10$ K  most likely points to the fact that the
electron density  decreases at these temperatures. The possible
reason of the decreasing is the transition of some part of
electrons from the well to the states of residual donors in the
buffer layer, to the states at the heterointerface, to  states
near the substrate/buffer boundary or near the surface. The
decreasing of $1/(eR_H)$ with the temperature increase is observed
for all the electron density in both structures investigated. The
explicit reason of the downturn of $1/(eR_H)$ remains unknown
therefore we restrict our analysis to the low temperature,
$T<9-10$ K.

The behavior of the second fitting parameter, which is  $\mu'$, is
shown in Figs.~\ref{F2}(c) and \ref{F2}(d). As seen, it increases
with the temperature increase for both cases, this increasing is
close to the linear one. The physical reason of such behavior will
be discussed below.

Now we are in position to consider the role of the {\it e-e}
interaction. There are different ways to extract experimentally the
{\it e-e} contribution. They follow from Eqs.~(\ref{eq06}) and
(\ref{eq07}) for the conductivity tensor components, and can be
outlined as follows: (i) the direct analysis of the magnetic field
dependence of $\sigma_{xx}$ and $\sigma_{xy}$; (ii) the analysis of
the parabolic-like negative magnetoresistance [see Eq.~(\ref{eq08})
below]; (iii) the analysis of the temperature dependencies of
$\sigma_{xx}$ and $\sigma_{xy}$ in high magnetic field, $B\gg
B_{tr}$, when the WL-correction is strongly suppressed; (iv) the
analysis of the temperature dependence of the Hall coefficient [see
Eq.~(\ref{eq12})]; (v) the analysis of the temperature dependence of
the conductivity at $B=0$.
\begin{figure}
\includegraphics[width=\linewidth,clip=true]{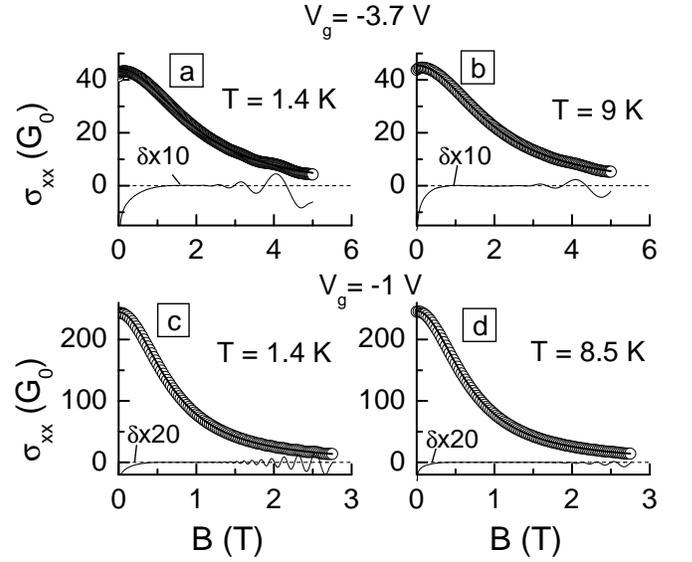}
\caption{(a--d) -- The magnetic field dependences of  $\sigma_{xx}$
for different temperatures and $V_g=-3.7$~V (a,b) and $V_g=-1$~V
(c,d), structure T1520. Symbols are the experimental data, solid
curves are the fit by Eq.~(\ref{eq06}). Curves labeled as
$\delta\times 10$ (a,b) and $\delta\times 20$ (c,d) are the
difference between experimental and fitting curves multiplied by a
factor $10$ or $20$, respectively. }
 \label{F4}
\end{figure}

Actually, if one firmly believes  that the {\it e-e} interaction is
the sole mechanism, which determines  the temperature and magnetic
field dependences of the conductivity, all the methods are not
independent and have to duplicate  each other. However, if there are
any additional mechanisms, for instance, temperature dependent
disorder, classical magnetoresistance, {\em etc}., the comparison of
the results of different methods gives a possibility to estimate the
role of ``additional'' mechanisms and is necessary for elucidating
of the contribution of {\it e-e} interaction more reliably. Let us
apply all listed methods in turn and analyze our experimental
results from this point of view.

\begin{figure}
\includegraphics[width=\linewidth,clip=true]{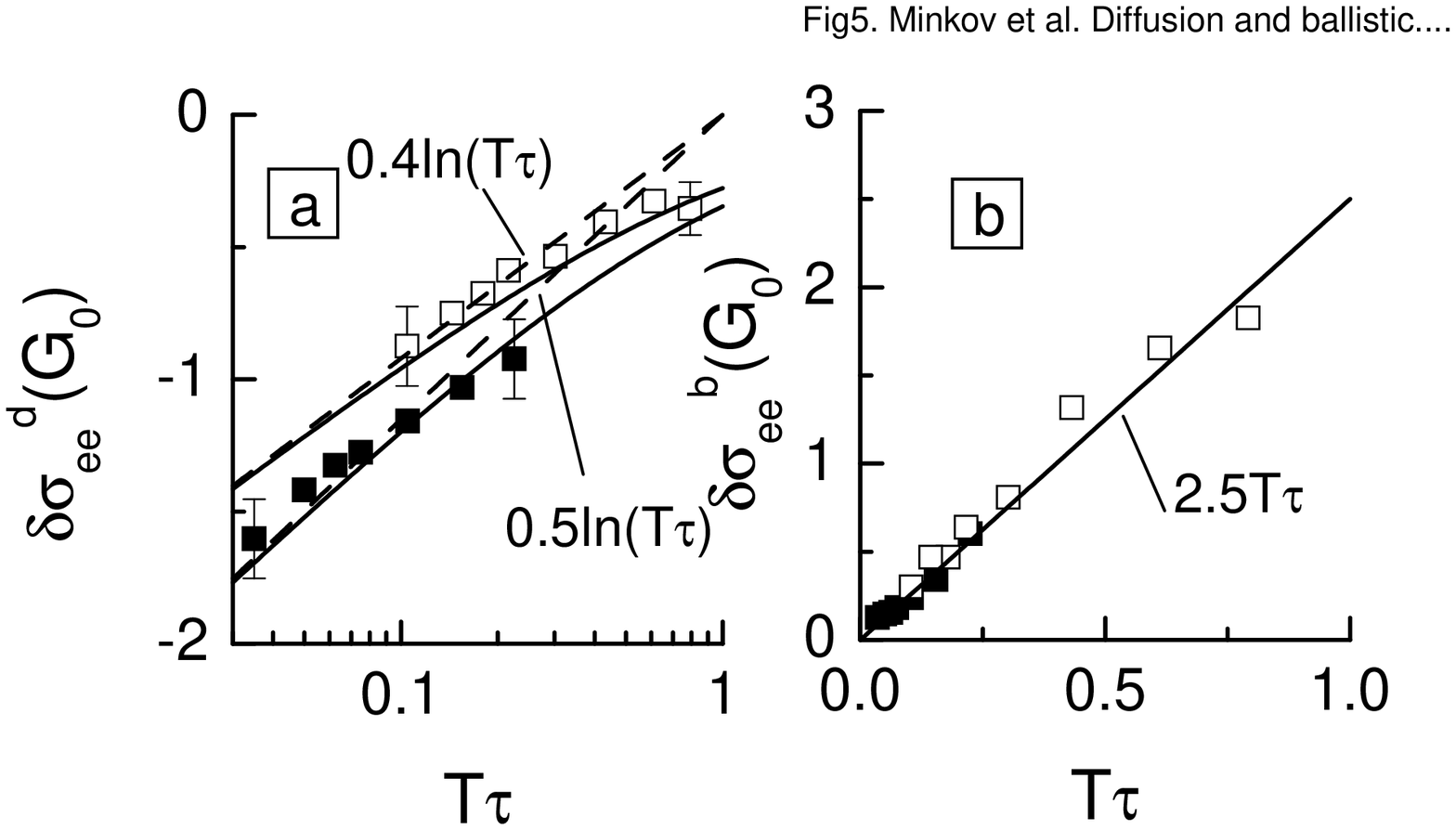}
\caption{The temperature dependences of the diffusion (a) and
ballistic (b) corrections. Symbols are the experimental data for
$V_g=-3.7$~V ($\blacksquare$) and $-1$~V ($\square$). Sold lines in
(e) are the improved expression, Eq.~(\ref{eq11}), for the diffusion
correction.}
 \label{F45}
\end{figure}

The first way of the finding of the {\it e-e} correction follows
from Eqs.~(\ref{eq06}) and (\ref{eq07}). One can fit the
experimental $\sigma_{xx}$-vs-$B$ curve  by Eq.~(\ref{eq06}) using
$\delta\sigma_{xx}^d$ and $\mu'$ as the fitting parameters, and $n$,
found from the fit of the experimental $\sigma_{xy}$-vs-$B$ curve
(see Fig.~\ref{F2}(a,b)). Figures~\ref{F4}(a--d) show the result of
the fitting procedure carried out in the magnetic field range from
$B=20\,B_{tr}$ to $B=1.5/\mu$. As seen, an agreement is excellent
for all gate voltages and  temperatures. The values of the diffusion
correction found by this way are presented in Fig.~\ref{F45}(a) as a
$\delta\sigma_{xx}^d$-vs-$T\tau$ plot. It is seen that the
temperature dependence of  $\delta\sigma_{xx}^d$ at $T\tau<0.4$ is
close to logarithmic $\delta\sigma_{xx}^d=K_{ee}G_0 \ln(T\tau)$ with
$K_{ee}\approx 0.5$ and $0.4$ for $V_g=-3.7$~V and $-1$~V,
respectively.

Note, $\delta\sigma_{xx}^d$ is obtained by this way as difference
between two large quantities  known with some error: the
experimental quantity $\sigma_{xx}$ and the quantity
$en\mu'/(1+\mu'^2B^2)$ [see Eq.~(\ref{eq06})], in which $n$ is also
experimental. To estimate this error we have fitted the data within
different magnetic field range. The equally good agreement between
experimental and calculated curves is observed in all the cases,
however the values of $\delta\sigma_{xx}^d$ is somewhat different
(see error bars in Fig.~\ref{F45}(a)).

The value of the second fitting parameter $\mu'$ found in this
procedure is presented in  Figs.~\ref{F2}(b) and \ref{F2}(d) in
which the $\mu'$-values found from the $\sigma_{xy}$-vs-$B$
dependence were shown. One can see that the fit of both
$\sigma_{xx}$ and $\sigma_{xy}$ components gives the
$\mu'$-values, which coincide with an accuracy of about $(1-2)$~\%
and demonstrate close to the linear temperature dependence.

\begin{figure}
\includegraphics[width=\linewidth,clip=true]{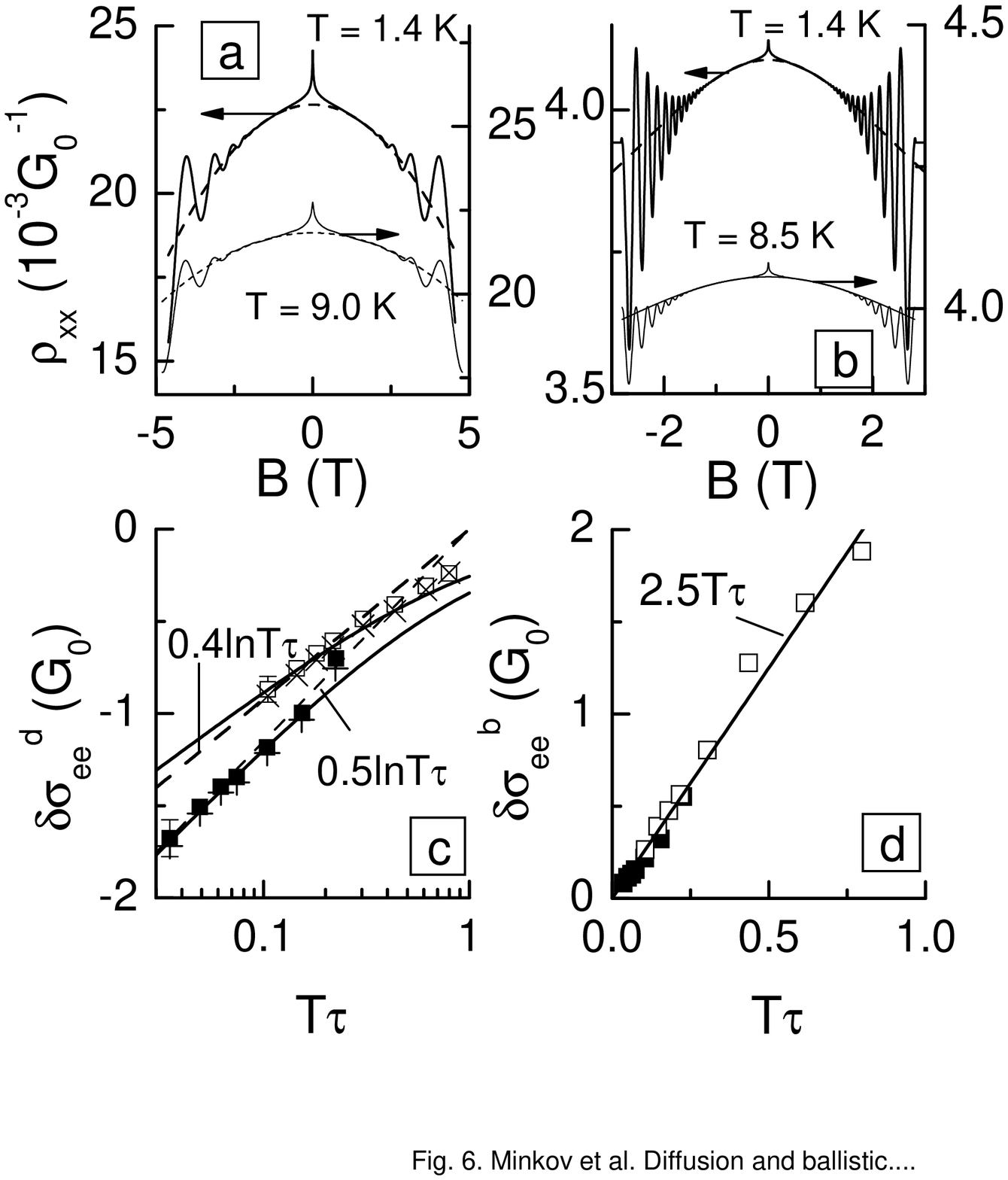}
\caption{(a) and (b) -- The parabolic-like magnetoresistance for
different temperatures for $V_g=-3.7$~V and $-1$~V, respectively,
structure T1520. Solid curves are the experimental data. Dashed
curves are the fit by Eq.~(\ref{eq08}). (c) and (d) -- The
temperature dependences of the diffusion and ballistic corrections,
respectively. Symbols are the experimental data for $V_g=-3.7$~V
($\blacksquare$, $+$) and $-1$~V ($\square$, $\times$) obtained by
the second  ($\blacksquare$, $\square$) and third ($+$, $\times$)
ways. Sold lines in (c) are the improved formula for the diffusion
{\it e-e} correction, Eq.~(\ref{eq11}). } \label{F6}
\end{figure}

What is the mechanism of the temperature dependence of $\mu'$? The
degeneracy of the electron gas  within the actual temperature range,
$T<10$~K,  remains strong: $E_F/T> 30$ for $V_g=-3.7$~V and $E_F/T>
55$ for $V_g=-1$~V. Therefore, the ionized impurities or roughness
scattering mechanisms cannot lead to  a mobility variation with
temperature. Obviously,  the phonon scattering is also not
responsible for such an effect, because  it has to lead to the
mobility decreasing with the temperature increase, which is opposite
to that observed experimentally for $\mu'$. So, we believe that the
temperature dependence of $\mu'$  results from the ballistic part of
the electron-electron interaction, which in the main really reduces
to renormalization of the mobility as we supposed writing
Eqs.~(\ref{eq06}) and (\ref{eq07}) out. The temperature dependence
of the  ballistic part  defined as $\delta\sigma_{ee}^b(T)=
en\,\delta\mu(T)$  with $\delta\mu(T)=\mu'(T)-\mu'(0)$ is presented
in Fig.~\ref{F45}(b). One can see that $\delta\sigma_{ee}^b$
linearly increases with temperature and all the experimental points
lie on the same straight line $\delta\sigma_{ee}^b(T)/G_0\simeq
2.5\,T\tau$ for both gate voltages.  It is important for the
following to note that the variation of $\delta\sigma_{ee}^b$ in our
temperature range is larger for $V_g=-1$~V than that for the case of
$V_g=-3.7$~V. It is sequence of the higher value of the transport
relaxation time $\tau$ in the first case (see Table~\ref{tab1}).

It should be mentioned, that the discrepancy between the
experimental $\sigma_{xx}$-vs-$B$ curves and calculated ones at
low magnetic field, resulting from the WL correction, continues up
to high magnetic field, $(15-20)\,B_{tr}$ [see Figs.~\ref{F4}(a)--
\ref{F4}(d)]. Therefore, if one uses the range of the magnetic
field including the lower fields in the fitting procedure, we can
obtain the wrong value of the correction.

The second way is based on the analysis of the parabolic-like
negative magnetoresistance. It directly follows from
Eqs.~(\ref{eq06}) and (\ref{eq07}) that the magnetoresistance should
have the form
\begin{equation}
\rho_{xx}(B,T)  \simeq
\frac{1}{en\mu'}-\frac{1}{(en\mu')^2}\left(1-\mu'^2 B^2\right)
\delta\sigma_{xx}^{d}(T). \label{eq08}
\end{equation}
Thus, fitting the experimental $\rho_{xx}$-versus-$B$ curve for a
given temperature by  Eq.~(\ref{eq08})  one can find both the
diffusion and ballistic corrections.  This method is free of
disadvantage of the previous one because $\delta\sigma_{xx}^{d}$ is
obtained not as a difference between two large values. As Figs.
\ref{F6}(a) and \ref{F6}(b) show Eq.~(\ref{eq08}) excellently
describes the experimental data. The temperature dependence of
$\delta\sigma_{ee}^d$ and $\delta\sigma_{ee}^b$ found from the fit
are presented in Figs.~\ref{F6}(c) and \ref{F6}(d). Comparison with
the results presented in Figs.~\ref{F45}(a) and \ref{F45}(b) shows a
good agreement with the data obtained by the first way.

The third  way is the analysis of the temperature dependence of
the Hall coefficient, $R_H=
\sigma_{xy}/[B(\sigma_{xy}^2+\sigma_{xx}^2)]$. It follows from
Eqs.(\ref{eq06}) and (\ref{eq07})  that the diffusion interaction
correction should be equal to
\begin{equation}
\delta\sigma_{xx}^d(T)=\frac{[R_H(T)-(en)^{-1}]en\mu'}{2R_H(T)}.
\label{eq12}
\end{equation}
To find $\delta\sigma_{xx}^d(T)$, we use the values of $n$ and
$\mu'$ obtained  from analysis of $\sigma_{xy}$-versus-$B$
dependences and the temperature dependence of $R_H$ [see
Figs.~\ref{F2}(a) and \ref{F2}(b)]. The results of such data
processing are plotted in Fig.~\ref{F6}(c) by crosses. One can see
that the value of the interaction correction and its temperature
dependence are very close to ones obtained with the use of the
previous methods. It should be stressed that the analysis of the
Hall coefficient gives the diffusion correction only.

\begin{figure}
\includegraphics[width=\linewidth,clip=true]{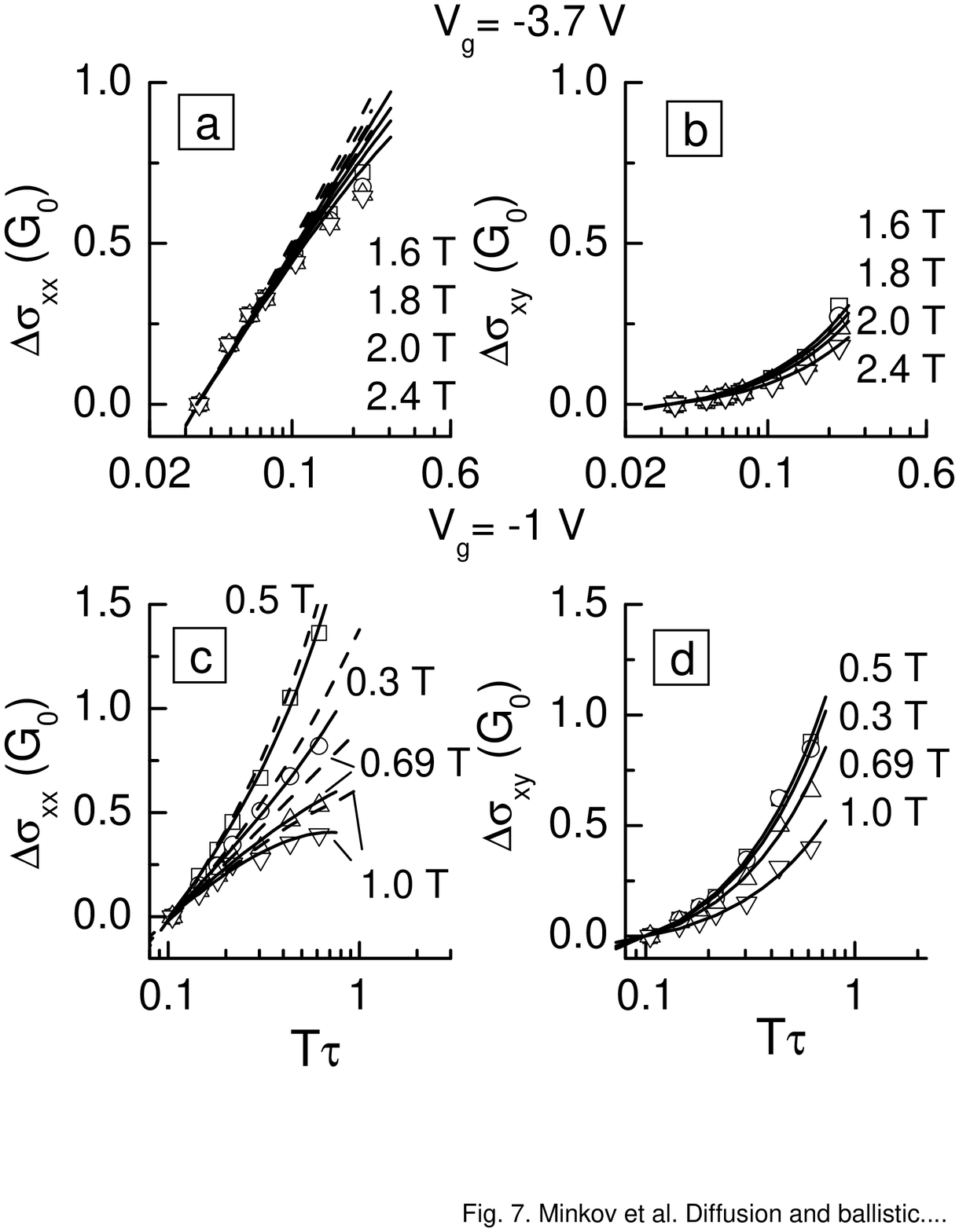}
\caption{The temperature  dependence of $\Delta\sigma_{xx}$ (a,c)
and $\Delta\sigma_{xy}$ (b,d) for different magnetic fields in
vicinity of $B=1/\mu$  for $V_g=-3.7$~V (a,b) and $V_g=-1$~V (c,d).
Symbols are the experimental results for $B=1.6$ ($\square$), $1.8$
($\circ$), $2.0$ ($\vartriangle$), and $2.4$~T ($\triangledown$)
(a,b), and $B=0.3$ ($\square$), $0.5$ ($\circ$), $0.69$
($\vartriangle$), and $1.0$~T ($\triangledown$) (c,d). Dashed lines
are Eqs.~(\ref{eq06}) and (\ref{eq07}) with $\mu'(T)$ presented in
Figs.~\ref{F2}(b) and \ref{F2}(d) and  $\delta\sigma_{xx}^d(T)$
given by Eq.~(\ref{eq05}) with $K_{ee}=0.45$  and $0.4$ for
$V_g=-3.7$~V and $-1$~V, respectively. Solid lines are obtained
analogously, but $\delta\sigma_{xx}^d(T)$ is calculated from the
improved formula, Eq.~(\ref{eq11}). Solid lines in panels (b) and
(d) coincide with dashed ones because $\delta\sigma_{xx}^d(T)$ does
not contribute to $\sigma_{xy}$.} \label{F7}
\end{figure}

The fourth way  is the analysis of the temperature dependence of
$\sigma_{xx}$ at high enough magnetic field where WL is suppressed.
For strictly diffusion regime the $\sigma_{xx}$-component should
logarithmically  depend on the temperature while $\sigma_{xy}$
should be temperature independent [see Eqs.~(\ref{eq03}) and
(\ref{eq04})]. In Fig.~\ref{F7} we have plotted the variation of
$\sigma_{xx}$ and $\sigma_{xy}$ with the temperature at different
magnetic fields (reference temperature is $T=1.4$~K). It is clearly
seen that  $\Delta\sigma_{xx}$ for  $V_g=-3.7$~V does not
practically depend on the magnetic field. At $T\tau\lesssim 0.1$ the
temperature dependence of $\Delta\sigma_{xx}$ is  logarithmic. The
slope is approximately equal to  $0.45$ that is very close to that
obtained with the help of above methods. The temperature dependence
of $\Delta\sigma_{xy}$ is significantly weaker and, as will be shown
below, results from the ballistic contribution via the mobility
renormalization.

The significantly different behavior is demonstrated by the data
obtained at $V_g=-1$ V. First of all, both  the
$\Delta\sigma_{xx}$-vs-$T$ and $\Delta\sigma_{xy}$-vs-$T$ plots
taken at different $B$ represent fan charts. Second, the variation
of $\sigma_{xy}$ with the temperature is comparable in magnitude
with that for $\sigma_{xx}$.  Obviously, both facts are a sequence
of the temperature dependence of the ballistic contribution, which
is larger in magnitude for this gate voltage due to higher value of
$T\tau$. Nevertheless, the diffusion contribution can be easily
extracted in this case as well. In framework of the model  used,
which reduces the ballistics to mobility renormalization, the
ballistic correction to $\sigma_{xx}$ is equal to zero when  $\mu
B=1$. Really, differentiating Eq.~(\ref{eq06}) with respect to $\mu$
one obtains:
\begin{equation}
\delta\sigma_{xx}^b=
\frac{\partial\sigma_{xx}}{\partial\mu}\delta\mu=\left.\frac{1-\mu^2B^2}{\left(1+\mu^2B^2\right)^2}\,en\,\delta\mu\right|_{\mu
B=1}=0. \label{eq09}
\end{equation}
That is why the temperature dependence of $\sigma_{xx}$ at $B=1/\mu$
should be wholly determined by the diffusion correction. Inspection
of  Fig.~\ref{F7}(c) shows that  the temperature dependence of
$\sigma_{xx}$ at $B=1/\mu=0.69$ T is actually close to logarithmic
up to $T\tau\simeq 0.25$ with the slope $0.4$, which coincides with
that found before [see Figs.~\ref{F45}(a) and \ref{F6}(c)].

Let us inspect how the model used describes the  temperature
dependences of $\Delta\sigma_{xx}$ and $\Delta\sigma_{xy}$ at $\mu
B\neq 1$. In Fig.~\ref{F7} we plot  the curves calculated from
Eqs.~(\ref{eq06}) and (\ref{eq07}) with $\mu'(T)$ presented in
Figs.~\ref{F2}(b) and \ref{F2}(d), and  $\delta\sigma_{xx}^d(T)$
given by Eq.~(\ref{eq05}) with $K_{ee}=0.45$  and $0.4$ for
$V_g=-3.7$~V and $-1$~V, respectively. One can see that our model
perfectly describes the data for $\Delta\sigma_{xy}$ [see
Figs.~\ref{F7}(b) and \ref{F7}(d)]. As for the temperature
dependence of $\Delta\sigma_{xx}$, there is satisfactory agreement
between the data and calculated results up to $T\tau\simeq 0.3$
[dashed lines in Figs.~\ref{F7}(a) and \ref{F7}(c)]. At higher
$T\tau$ values a discrepancy between calculated curves is evident,
the stronger magnetic field the more pronounced discrepancy is. In
the next section we propose an improvement of Eq.~(\ref{eq05}) for
the diffusion contribution, which gives an excellent accordance over
the whole $T\tau$-range.
\begin{figure}
\includegraphics[width=\linewidth,clip=true]{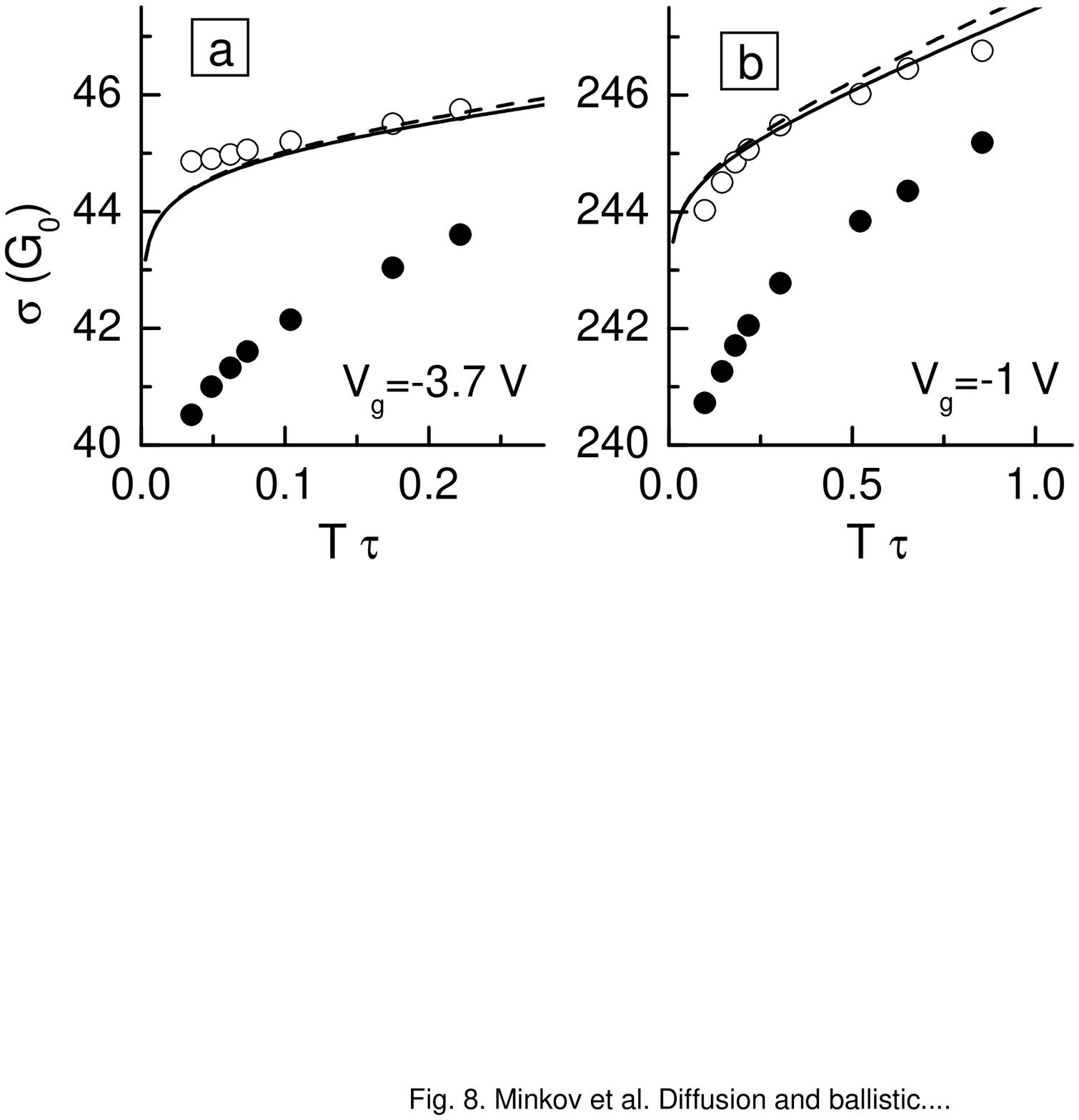}
\caption{The temperature dependences of the conductivity at $B=0$.
Solid symbols are  for the conductivity measured experimentally.
Open symbols are the same data after subtraction of the interference
quantum correction. Lines are drown as described in the text.}
 \label{F8}
\end{figure}

\begin{figure}[b]
\includegraphics[width=\linewidth,clip=true]{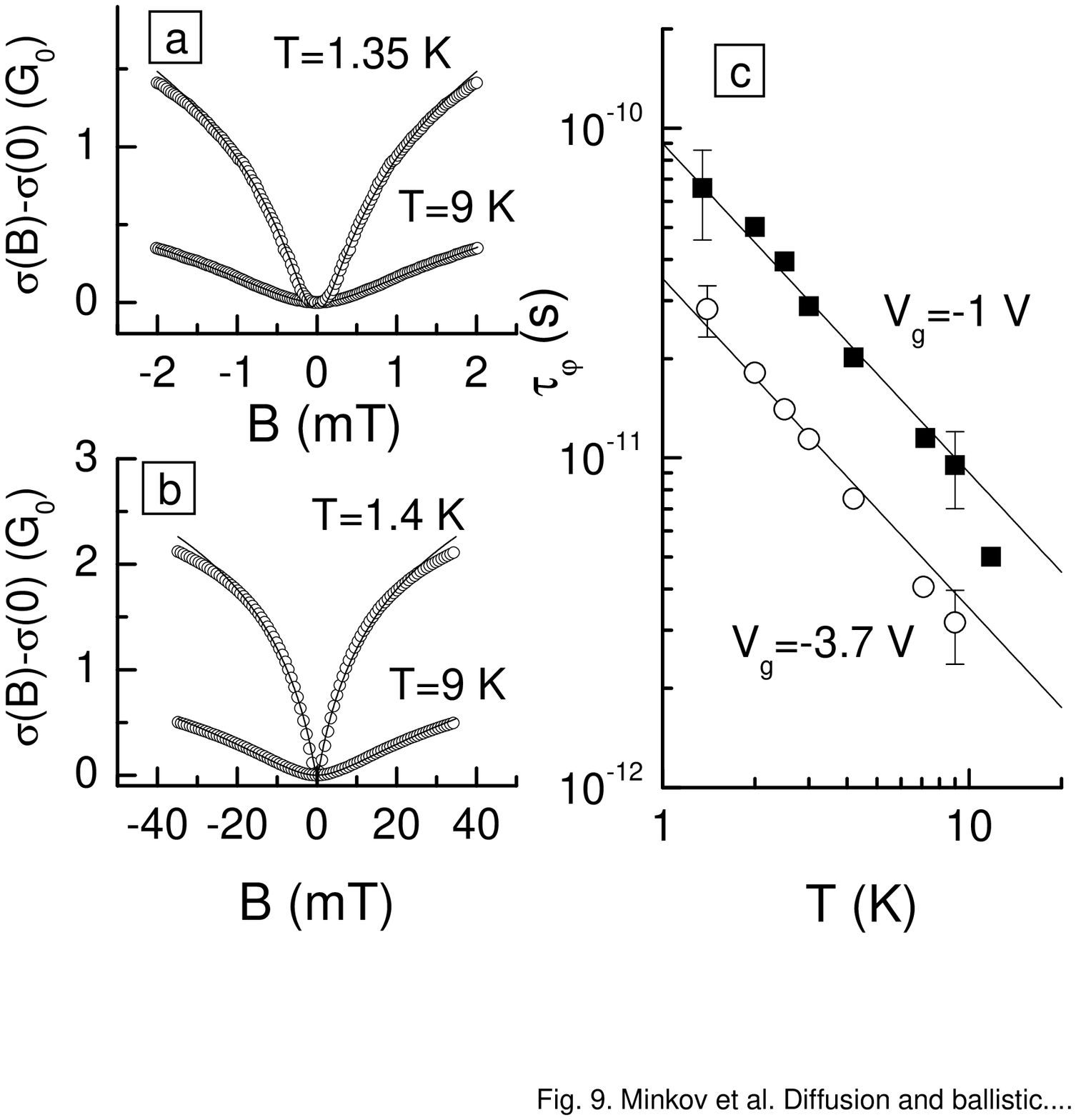}
\caption{(a, b) -- The magnetoconductivity caused by suppression of
the interference quantum correction measured at  two temperatures
and two gate voltages. Symbols are the experimental data. Curves are
the best fit by formulae from Ref.~\onlinecite{Hik80,Wit87} with
$\tau_\phi$ presented for different temperatures in panel (c). Lines
in (c) are $T^{-1}$-law.}
 \label{F9}
\end{figure}

Up to now we determined the interaction correction in the presence
of a magnetic field. Let us now turn to the last method and consider
the correction at $B=0$. The experimental temperature dependences of
the conductivity at $B=0$ are presented in Fig.~\ref{F8} by solid
circles. Since this dependence is determined by both the WL and
interaction correction, one should exclude the WL contribution to
find the interaction contribution. For this purpose  we have
measured the low-field magnetoresistance [Figs.~\ref{F9}(a) and
\ref{F9}(b)] caused by suppression of the WL correction. Analyzing
the shape of magnetoresistance curves using standard
procedure\cite{Hik80,Wit87} we have found the phase relaxation time
($\tau_\phi$) and its temperature dependence [Fig.~\ref{F9}(c)].
After that the WL quantum correction has been calculated according
to Eq.~(\ref{eq011}) and subtracted from the experimental values of
conductivity at $B=0$. Thus, we have obtained the conductivity,
which temperature dependence is caused only by the interaction
corrections [shown by open symbols in Figs.~\ref{F8}(a) and
\ref{F8}(b)]. To compare these data with the results obtained above,
we have calculated $T$-dependences of
$en\mu(0)+\delta\sigma_{ee}^d+\delta\sigma_{ee}^b$ using $\mu(0)$
from Table~\ref{tab1}, $\delta\sigma_{ee}^b=2.5\,G_0\,T\tau$, and
$\delta\sigma_{ee}^d(T)=0.45\,G_0\,\ln(T\tau)$ and
$0.4\,G_0\,\ln(T\tau)$ for $V_g=-3.7$~V and $-1$~V, respectively.
The results are shown in Fig.~\ref{F8} by  dashed lines. So, the
parameters obtained in the presence of magnetic field well describe
the temperature dependence of the conductivity at zero magnetic
field.

Thus, we have found the diffusion and ballistic contributions of the
interaction correction to the conductivity by the different ways.
The fact that all these methods give close results shows that other
``parasitic'' mechanisms, which could contribute to the temperature
and magnetic field dependences, are negligible.
We turn now to  discussion.

\section{Discussion}

\begin{figure}
\includegraphics[width=0.8\linewidth,clip=true]{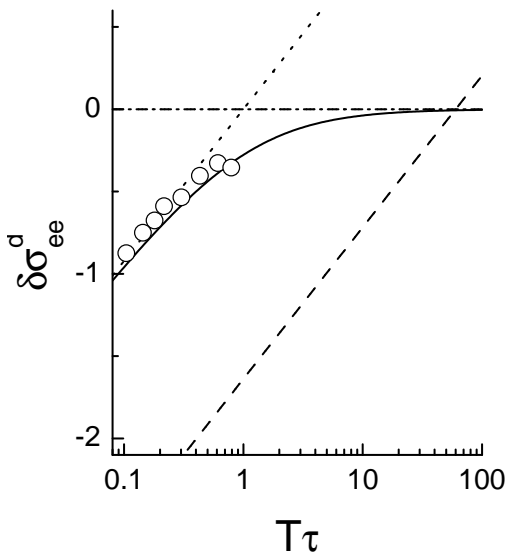}
\caption{The $T\tau$-dependence of $\delta\sigma^d_{ee}$ for
$V_g=-1$~V. Symbols are the experimental results. Dotted and dashed
lines are dependences  $K_{ee}\, G_0 \ln{T\tau}$ and $K_{ee}\, G_0
[\ln{T\tau}-\ln{(k_Fl/2)}]$, respectively. Solid line is the
dependence $-K_{ee}\, G_0 \ln{\left[1/(T\tau)+1\right]}$,
Eq.~(\ref{eq11}). In all the cases $K_{ee}=0.4$. }
 \label{F11}
\end{figure}

First of all, let us consider the absolute value of the diffusion
part of the interaction correction. As noted in Section~\ref{sec:TB}
the authors of Ref.~\onlinecite{Zala01} ``have chosen the argument
of the logarithm in Eq.~(\ref{eq02}) to be $E_F/T$ instead of the
usual $1/T\tau$ to emphasize that contrary to the naive expectations
the logarithmic term persists up to temperatures much larger than
$1/\tau$''. It means that the logarithmic part of the correction can
be written as $-K_{ee}G_0[\ln(k_F l/2)- \ln(T\tau)]$. The question
is: does temperature independent term $-K_{ee}G_0\ln(k_F l/2)$
contribute to $\delta\sigma_{xx}^d$ or not? To clarify we have
plotted in Fig.~\ref{F11} the both theoretical $T\tau$-dependences
[Eq.~(\ref{eq05}) and logarithmic part of Eq.~(\ref{eq02})], using
$K_{ee}$ found from the temperature dependence of $\sigma_{xx}$
(forth method) and parameters from Table~\ref{tab1}.  In the same
figure we present the experimental data for $V_g=-1$~V when the
Drude conductivity is maximal in magnitude. Mere  it is not needed
to involve the $-K_{ee}G_0\ln(k_F l/2)$-term to describe the
experiment.  Any temperature independent contribution that might
exist in $\sigma_{xx}^d$  is lower than $(0.1-0.2) G_0$.

Thus, two parts of the logarithmic correction are different. In the
presence of a magnetic field, the first one, $K_{ee}G_0\ln(T\tau)$,
contributes only to $\sigma_{xx}$ but does not to $\sigma_{xy}$.
Just this term we figure out experimentally. The second term,
$-K_{ee}G_0\ln(k_F l/2)$, contributes both to $\sigma_{xx}$ and to
$\sigma_{xy}$ and, in fact, reduces to the renormalization of the
transport relaxation time, i.e.,  the mobility.

Next issue, which should be pointed out is the parabolic-like
negative magnetoresistance in the high magnetic field.
Fig.~\ref{F6}(a,b) shows that at low $T\tau$ value such a behavior
is observed against a background of the Shubnikov-de Haas
oscillations far exceeding the magnetic field $B=1/\mu$. However, at
large $T\tau$-value  the monotonic part of the experimental curve
runs noticeably steeper at $B\gtrsim 2/\mu$ [see lower curves in
Fig.~\ref{F6}(b)]. From our point of view, it can be resulted from
the {\it e-e} interaction as well. As shown in
Ref.~\onlinecite{Gor04} in presence of long-range potential the
ballistic contribution is suppressed at $B=0$ and restores at high
magnetic field. Because in our structures the ballistic correction
is positive, it is equivalent  to the mobility increase  that, in
its turn, leads to additional decreasing of $\rho_{xx}$ with
$B$-increase at $B>1/\mu$. This effect is proportional to the
ballistic contribution and therefore reveals itself at $V_g=-1$~V
when the value of $T\tau$ is larger. The analogous deviation of
$\rho_{xx}$ from parabola was observed in Ref.~\onlinecite{Li03}.
The interpretation was also based on the magnetic field dependence
of the ballistic part of the {\it e-e} interaction correction.

Let us finally discuss  the diffusion correction at large $T\tau$.
As seen from Figs. \ref{F45}(a), \ref{F6}(c) at $T\tau\lesssim
0.2-0.3$ its temperature dependence is close to the logarithmic one.
However at larger $T\tau$ the systematic deviation down is evident.
Besides, the temperature dependence of $\sigma_{xx}$ at $T\tau >
0.2-0.3$ for different $B$ is described in the framework of the
model used only qualitatively [see Fig.~\ref{F7}(a,c)]. We have
found that the agreement can be  improved if one replaces the
argument $1/(T\tau)$ in logarithm in Eq.~(\ref{eq05}) by
$1/(T\tau)+1$ that removes the divergence of diffusion contribution
with $T\tau$-increase:
\begin{eqnarray}
{\delta \sigma_{xx}^{d}(T)\over G_0}&=&-\left[1+3\left(1-
\frac{\ln(1+F_0^\sigma)}{F_0^\sigma}\right)\right]\ln{\left(\frac{1}{T\tau}+1\right)}\nonumber
\\
&\equiv&-K_{ee}\ln{\left(\frac{1}{T\tau}+1\right)}, \label{eq11}
\end{eqnarray}
After  such  modification the agreement with the experimental data
becomes excellent within whole $T\tau$ range as Figs.~\ref{F45}(a),
\ref{F6}(c), \ref{F7}, and \ref{F8} demonstrate. It should be
mentioned that Eq.~(\ref{eq11}) in combination with
Eqs.~(\ref{eq06}) and (\ref{eq07})  reproduces the $1/T$ temperature
dependence of ballistic asymptotics of $\delta\rho_{xx}(T)/B^2$ and
$\delta R_H(T)/B$, in a qualitative agreement with
Refs.~\onlinecite{Gor04} and \onlinecite{Zala01-1}, respectively.
The numerical coefficients in front of these asymptotics depend on
details of disorder and were calculated in Refs.~\onlinecite{Gor04}
and \onlinecite{Zala01-1} only for the case of a purely white-noise
disorder and thus should not be necessarily reproduced in
experiments on realistic structures.

\begin{figure}
\includegraphics[width=0.8\linewidth,clip=true]{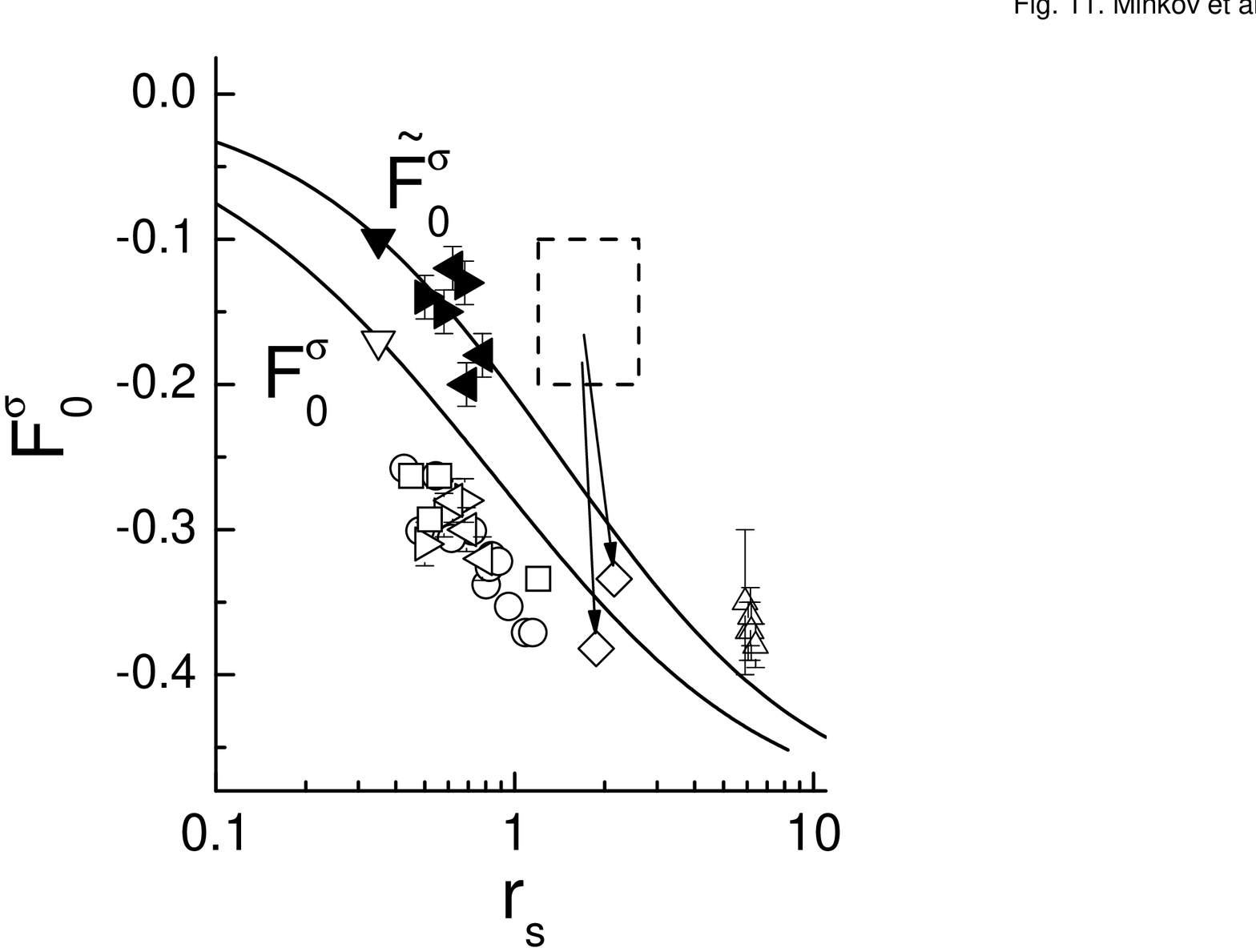}
\caption{The $r_s$-dependence of the Fermi-liquid constants
$F_0^\sigma$ and  $\widetilde{F}_0^\sigma$. Lines are calculated
according to Ref.~\onlinecite{Zala01}. Symbols are the experimental
results for structures T1520 ($\blacktriangleright,\triangleright$)
and 3510 ($\blacktriangleleft,\triangleleft$), for the samples with
$k_F l>5$ from Ref.~\onlinecite{Min03} ($\circ$), structures from
Ref.~\onlinecite{Min01} ($\square$), Ref.~\onlinecite{Yasin04}
($\triangle$), and Ref.~\onlinecite{Renard05}
($\triangledown,\blacktriangledown$). Dashed box indicates an
approximate range of $r_s$ and $F_0^\sigma$ for the structures from
Ref.~\onlinecite{Gal04,Li03}. ($\diamond$) -- The new $F_0^\sigma$
values for two samples from Ref.~\onlinecite{Li03} obtained in our
interpretation (see the text for details). Open and solid symbols
are for $F_0^\sigma$ and $\widetilde{F}_0^\sigma$, respectively.}
 \label{F12}
\end{figure}

Analogous measurements were performed for the structure 3510, where
the electron density and mobility were controlled by illumination
due to persistent photoconductivity (see Table~\ref{tab1}). The
parameter $T\tau$ for this structure laid within the interval from
$0.07$ to $0.7$. All five ways of determination of the interaction
correction described above give consistent results also.  It turns
out that the values and the temperature dependences both of the
diffusion and ballistic corrections are very close to that for
structure T1520.

In the framework of theory\cite{Zala01} the interaction corrections
are governed  by the Fermi-liquid parameter $F_0^\sigma$ for the
diffusion correction and by $\widetilde{F}_0^\sigma$ for the
ballistic one [see Eq.~(\ref{eq02})], these parameters depend on
$r_s$ only.  The values of $F_0^\sigma$  and
$\widetilde{F}_0^\sigma$ obtained for both structures investigated
in this paper and $F_0^\sigma$ found in our previous
papers\cite{Min01,Min03} are shown in Fig.~\ref{F12}.\footnote{As
shown in Ref.~\onlinecite{Min03}, the experimental value of
$F_0^\sigma$ depends not only on the gas parameter $r_s$, as it
should be theoretically in the case of $k_Fl\gg 1$, but on the
disorder strength as well. Therefore, in Fig.~\ref{F12} we present
the data from Ref.~\onlinecite{Min03} relating to the relatively
high value of $k_Fl$: $k_Fl>5$.} One can see that all the data
correlate well. The $r_s$-dependences of both $F_0^\sigma$ and
$\widetilde{F}_0^\sigma$  are close to theoretical ones, though the
experimental points for $F_0^\sigma$ fall systematically below the
corresponding theoretical curve.

Let us compare our results  with that obtained by other authors for
the analogous 2D electron systems.  Recently, the paper by Renard
{\it et al} \cite{Renard05} devoted to an experimental study  of
very-low-mobility GaAs quantum wells  in a temperature range $1.5 -
110$~K has been released. The value of the parameter $T\tau$ in
these systems was less than $0.3$ even at the highest temperature.
So, only the beginning of the crossover from the diffusive to the
ballistic regime is spanned in this paper. The gas parameter $r_s$
in samples investigated was equal to $0.3-0.35$. The authors were
able to describe the longitudinal conductivity and the Hall effect
within framework of the theories\cite{Zala01,Zala01-1} using the
theoretical values of $F_0^\sigma$ and $\widetilde{F}_0^\sigma$ from
Ref.~\onlinecite{Zala01} (shown by $\triangledown$ and
$\blacktriangledown$ in Fig.~\ref{F12}). Some difference in the
interpretation of the data in Ref.~\onlinecite{Renard05} with
respect to our analysis is result of the fact that the range of the
magnetic field in this paper was limited by the value of about $8\,
B_{tr}$. Figs.~\ref{F4}(a--d) show that the interference correction
under this condition is not completely suppressed. We suppose that
neglect of this fact gives some error in determination of the value
of the {\it e-e} correction.

The {\it e-e} interaction correction in GaAs 2D systems of high
quality with extremely low electron density was systematically
studied in Ref.~\onlinecite{Yasin04}. It has been shown that the
theory\cite{Zala01} consistently describes the temperature
dependences of the conductivity in zero-magnetic field and of the
Hall resistivity in different  magnetic fields.   The parameters
$F_0^\sigma$ extracted from $\sigma(T)$ and $\rho_{xy}(T)$ are close
to each other. As seen from Fig.~\ref{F12} the data from this paper
shown by open triangles correlate well with our results despite the
large $r_s$ value.

The role of this correction within wide $T\tau$ range was studied
also in the papers by Galaktionov {\it et al}\cite{Gal04} and Li
{\it et al}.\cite{Li03} They investigated  GaAs heterostructures
with the relatively high electron mobility at $T\tau\simeq
0.03-0.3$. The authors presumed that the scattering was governed by
the long-range potential and therefore applied the theory by Gornyi
and Mirlin.\cite{Gor03,Gor04} The values of the Fermi-liquid
parameter obtained in this paper (indicated in Fig.~\ref{F12} by
dashed box) strongly differ from our and all other results. It
should be noted that the authors restricted themselves by
consideration of $\rho_{xx}(B)$ and has not analyzed other effects.
If one reinterprets these data supposing that the white-noise
disorder is the main scattering potential,  the parabolic-like
negative magnetoresistance within the framework of our model should
be determined by the diffusion correction only. Retreating the data
presented in Fig.~1 from Ref.~\onlinecite{Li03} for two samples with
$n=6.8\times10^{10}$ and $9\times10^{10}$~cm$^{-2}$ in such a
manner, we obtain the new values of $F_0^\sigma$ (diamonds in
Fig.~\ref{F12}),  which accord well with the other results.

\section{Conclusion}

We have experimentally studied the electron-electron interaction
correction to the conductivity of two-dimensional electron gas in
Al$_x$Ga$_{1-x}$As/GaAs/Al$_x$Ga$_{1-x}$As and GaAs/In$_x$
Ga$_{1-x}$As/GaAs  single-quantum-well heterostructures in a wide
range of $T\tau$-parameter, $T\tau=0.03-0.8$, covering the diffusion
and ballistic regimes. We have shown that the correction is
separated into two parts, which are distinguished by the manner how
they modify the conductivity tensor in the presence of a magnetic
field. The first part, or the diffusion correction, contributes to
$\sigma_{xx}$ only. The contribution to $\sigma_{xy}$ is equal zero.
The experimental value and the temperature dependence of the
diffusion correction is proportional to
$\ln\left[1/(T\tau)+1\right]$, Eq.~(\ref{eq11}). We have shown that
this part does not include the temperature independent term
$-K_{ee}G_0\ln(E_F\tau)$.  The second part of the interaction
correction, the ballistic part, is reduced to the renormalization of
the transport relaxation time $\tau$, that results in appearance of
the temperature dependence of the mobility. The ballistic correction
linearly increases with the temperature increase.  This model allows
us to describe consistently  the behavior of the components  both of
the resistivity and of the conductivity tensors with magnetic field
and temperature as well as the temperature dependence of the
conductivity without magnetic field.  We have experimentally
determined the values of the Fermi-liquid parameters $F_0^\sigma$
and $\widetilde{F}_0^\sigma$ and found them to be close to those
predicted theoretically.

\subsection*{Acknowledgments}
We thank Igor Gornyi for useful discussion. This work was supported
in part by the RFBR (Grants 03-02-16150, 04-02-16626, and
05-02-16413), the CRDF (Grants EK-005-X1 and Y1-P-05-11), and the
INTAS (Grant 1B290).


\end{document}